\documentclass[preprint,showpacs,preprintnumbers,amsmath,amssymb,aps,altaffilletter,nobibnotes]{revtex4}

\usepackage{graphicx,calc}
\usepackage{bm}

\begin{document}
\title{Thermal rectifier from deformed carbon nanohorns}

\author{Gang Wu}
\email[electronic mail:]{wugaxp@gmail.com} \affiliation{Department
of Physics and Centre for Computational Science and Engineering,
National University of Singapore, Singapore 117542-76, Republic of
Singapore }

\author {Baowen Li}
\email[electronic mail: ]{phylibw@nus.edu.sg}
 \affiliation{Department of
Physics and Centre for Computational Science and Engineering,
National University of Singapore, Singapore 117542-76, Republic of
Singapore\\
NUS Graduate School for Integrative Sciences and Engineering,
Singapore 117597, Republic of Singapore }

\begin{abstract}
We study thermal rectification in single-walled carbon nanohorns
(SWNHs) by using non-equilibrium molecular dynamics (MD) method. It
is found that the horns with the bigger top angles show larger
asymmetric heat transport due to the larger structural gradient
distribution. This kind of gradient behavior can be further adjusted
by applying external strain on the SWNHs. After being carefully
elongated along the axial direction, the thermal rectification in
the elongated SWNHs can become more obvious than that in undeformed
ones. The maximum rectification efficiency of SWNHs is much bigger
than that of carbon nanotube intramolecular junctions.
\end{abstract}


\date{\today}
\maketitle

Although human being has mastered the techniques to control the
electronic flux by using elements such as diodes and/or
transistors for a very long time, how to fabricate such devices to
control the heat flux is still an open and challenging question.
Given the importance of heat to human being, thermal controlling
devices should also play an important role in our future life,
ranging from nanoscale calorimeters, cooling of microelectronic
processors, to macroscopic refrigerators and energy-saving
buildings.

The extensive study of heat conduction in low dimensional systems
from microscopic point of view  in past two decades\cite{r3, r3_2,
r3_3} has enriched our understanding of the heat conduction in
finite size systems. In turn, some interesting inventions have been
made recently. For example, two interesting solid state thermal
diode/rectifier models have been proposed from nonlinear
lattices.\cite{r4, r5, r5_2} Based on the {\it negative differential
thermal resistance} (NDTR), a thermal transistor model has been
constructed.\cite{r6} By combining thermal transistors in different
ways, elementary thermal logic gates such as NOT, AND/OR gates have
been also demonstrated recently.\cite{Wang-Li2007} More
significantly, the two segment model of thermal rectifier proposed
in Ref \onlinecite{r4, r5} has been experimentally realized by using
gradual mass-loaded carbon and boron nitride nanotubes.\cite{r7} It
has been illustrated that the carbon nanotube intramolecular
junctions also exhibits thermal rectifying effect.\cite{WG-LB2007}
However, the maximum rectification efficiency of the real materials
seems to be no more than $1.5$. Therefore, to find more efficient
thermal rectifier is becoming of primary importance for future
application.

Nano carbon materials,\cite{r8} such as carbon nanohorn
aggregates,\cite{r9} have recently attracted considerable
attention since they are promising nanoscale materials with
potential applications in broad fields, such as catalyst supports
in fuel cell electrodes,\cite{r10} in gas storage
devices,\cite{r12, r13} as well as in drug delivery
systems.\cite{r14}

The nanohorn is composed of a horn-shaped, cap-closed carbon
nanotube (CNT) formed out of a single-walled graphene sheet and
several of them are put together to form a flowerlike aggregate
with a rather uniform diameter of about 100 nm. CNHs have very
large surface areas which enable them to support very fine
catalyst particles and to entrap drug molecules and even gases.

If the CNH is made of one single graphene sheet, we call it
single-walled carbon nanohorn (SWNH). Two examples are shown in Fig.
1.  The SWNH has an angle of about 19\r{ }, a diameter of 2-4 nm,
and a length of about 50 nm.\cite{r9} They come together to form a
spherical aggregate, and at their center the SWNHs coalesce (see
Fig. 6 of Ref. \onlinecite{r8}).

In this paper, we will demonstrate that SWNH can be a very efficient
thermal rectifier. Moreover, we find that the force constants along
the SWNHs' axes can be adjusted in the elongated structure. With
suitable tensile stress, the rectification in deformed SWNH becomes
more obvious than that in undeformed one, which makes the deformed
SWNH an ideal candidate for thermal rectifier.

The SWNHs are obtained by cutting out 60\r{ } and 120\r{ } sectors
from a graphene sheet and then gluing together the two cut sides of
the sheet. They are illustrated in Fig. 1 and named as SWNH-60 and
SWNH-120 in the following text. In addition, we also perform
simulation on a smaller SWNH-120 structure, the central axial length
of which is only half of the original SWNH-120. This smaller
structure is named SWNH-120-SEG. In Fig. 1, the blue atoms are fixed
in MD process, and orange atoms are put in thermostats, which are
realized by the Nos\'e-Hoover thermostat.\cite{r15_2} The
temperatures of left and right ends are $T_L $ and $T_R $,
respectively. When $T_L \ne T_R $, temperature gradient can be built
up in the  SWNH. For convenience, we introduce two variables:
$\Delta T = \frac{T_L - T_R }{2\left\langle T \right\rangle }$, and
$\left\langle T \right\rangle = \frac{T_L + T_R }{2}$. In this work,
$\left\langle T \right\rangle $ is always kept at 290 K. The heat
flux running from the narrow end to the wide one is defined as
positive.

The velocity Verlet method is adopted to integrate the equations of motions
with the time step of 0.51 fs. The typical total MD process is $1\times 10^7$
steps which is about 5.1 ns, and the statistic averages of interesting
quantities start from half of the MD process, i.e., $5\times 10^6$ steps are
used to relax the system to stationary state.

The instant temperature of atom $i$ is defined as $T_i \left( t
\right) = \frac{m_i }{3k_B }\left( {v_x \left( t \right)^2 + v_y
\left( t \right)^2 + v_z \left( t \right)^2} \right)$, where $m_i $
is the mass, $v\left( t \right)$ is the velocity at time $t$ and
$k_B $ the Boltzmann constant. This is a result of energy
equipartition theorem.

The thermal flux is obtained from the thermostats, i.e., the total
work from the thermostats can be regarded as the heat flux runs from
thermostats to the system. For more details, please refer to our
recent work.\cite{WG-LB2007} The second-generation reactive
empirical bond order potential \cite{r16} is employed to describe
the C-C interactions.

We use quantity $L / L_0 $ to describe the strength of tensile
strain, where $L$ and $L_0 $ are the central axial length after and
before elongation, respectively. $L_0 $ is obtained after fully
structural optimization.

The \emph{rectification},  which was defined as heat flux ratio $|J_
n / J_ p| = - J_ n / J_ p $,\cite{r4,r5} is used to quantify the
asymmetric heat conduction. Where $J_ n$ means the heat flux runs
from wide end to the narrow one, and $J_ p$ means the heat flux runs
in the opposite direction.

We first study the heat conduction of SWNHs with different top
angles under different temperature difference ($\Delta T = $-0.05,
0.05, -0.30 and 0.30) and/or tensile strains. The results are
shown in Fig. 2.

It is obvious that $|J_n| > |J_p|$ in most cases. That's to say, the
heat flux runs preferentially along the direction of radius and mass
decreasing. This result bears an analogy to the recent experimental
one,\cite{r7} which implies that radius and mass gradients are
possible reasons of thermal rectification.

In order to understand the origin of the thermal rectification, we
can consider the similar phenomena in carbon nanotube intramolecular
junctions (IMJs) \cite{WG-LB2007} as a reference. Both of the SWNHs
and IMJs can be considered as curved graphene surfaces, so their
phonon spectra are similar to that of the graphene, with some
difference induced by the curvature effect and Born-von Karman
boundary condition along the circumferential direction. For the
narrow end of the SWNHs, the curvature effect is much significant
and the phonon spectra deviates much from that of the graphene;
whereas for the wide end, the phonon spectra approaches to that of
the graphene. Therefore, the phonon spectra along the central axis
of the SWNH changes gradually from narrow to wide end. As we have
known, the matching and/or mismatching of the energy spectra along
direction of the heat transport is the underlying mechanism of the
rectification in nonlinear lattice systems \cite{r4, r5, r5_2} and
IMJs,\cite{WG-LB2007} we can easily understand the appearance of the
thermal rectification in SWNHs. Furthermore, it was shown that heat
can more easily transport from wider side to narrower one in IMJs.
This behavior is also consistent with our observation of the thermal
rectification in SWNHs. This correspondence also implies the common
underlying mechanism of the thermal rectification in both systems.

The SWNH-120-SEG can be elongated more than both SWNH-60 and
SWNH-120. In Fig. 2, the maximum possible tensile strain for
SWNH-120-SEG is about 1.06, while that for SWNH-60 and SWNH-120 is
about 1.03. This is because the largest strain occurs at the narrow
end, when system is elongated, the narrower end of longer SWNH will
endure more strain than that of shorter ones due to the higher
structural asymmetry.

It is found from the inset of Fig. 2 that both the temperature
difference ($\left| {\Delta T} \right|$) and the tensile strain
($L/L_0$) have large influence on the heat flux $J$, but their
effects are in contrast with each other. Namely, an increase in
temperature gradient will cause an increase of heat flux, but when
tensile strain becomes stronger, heat flux $J$ decreases. When the
system is elongated, the interatomic force constants decreases,
which leads to the decrease of the heat flux.

In addition, more interesting are the influences of the temperature
difference and the tensile strain on the thermal rectification.

We find that an increase in temperature difference $\left| {\Delta
T} \right|$ will result in an increase in rectification. The
rectification is found to be mainly controlled by the
middle-frequency optical phonon modes,\cite{WG-LB2007} which can be
further excited when temperature gradient increases. In fact, these
middle-frequency optical phonon modes are most sensitive for
curvature effect in CNT,\cite{r17_2} so their participation in
transport process also play most important role in the thermal
rectification of SWNHs. In other word, these phonon modes of
narrower end is sensitive to temperature change, which makes thermal
rectification changes when temperature gradient changes.

Under the same $\left| {\Delta T} \right|$, SWNH-60 shows smallest
rectification when the strain is weak. This is reasonable because
the top angle $\theta $ of the horn can be regarded as an indicator
of the structural asymmetry and curvature gradient along central
axis. When $\theta = 0$ is considered as an extreme situation, the
SWNH will degenerate to SWNT and phonon spectra gradient disappears,
and as we known, there does not exist thermal rectification in SWNT.
Moreover, the rectification in SWNH-120 is even larger than that in
SWNH-120-SEG. This is due to the larger structural difference
between the two ends of SWNH-120. Therefore, if we want to design a
device exhibiting large rectification, the best choice is to find a
system with large structural asymmetry. On the other hand,
considering the mass density along the $Z$ direction is not
homogenous, the SWNHs can be also thought as a mass graded system.
Recent study \cite{r18} has demonstrated that the thermal
rectification exists in mass graded anharmonic system. So the mass
gradient system, which is a special case of structural asymmetry
system, is hopefully to be a thermal rectifier. In fact, recent
experiment of thermal rectifier \cite{r7} has proven this idea.

As shown in Fig. 2, when tensile strain becomes stronger, the
rectification does not change monotonically. However, the trend of
the rectification versus strain seems similar for all the horns.
Under weak strain, the rectification increases with the increasing
strain; nevertheless, the rectification will decrease if the strain
becomes large. We can understand this as following qualitatively.

When the system is elongated, the interatomic force constants do not
decrease in the same rate along the $Z$ direction. Actually, the end
with large radius can hardly be elongated. As a result, high local
strain occurs near the narrow end and the force constants along the
$Z$ direction will change gradually. Consequently, the nanohorn
becomes a force constant graded system. Therefore, the asymmetry in
curvature and mass distribution needs to compete (or cooperate) with
the asymmetry in force constant distribution. Furthermore, the force
constants are not linearly dependence on the bond length when the
strain is strong, so the gradient in force constants does not always
increase with the increase of strain. Above multifold reasons
finally lead to complicated change of the phonon spectra gradient,
and in turn to the complicated trend of the rectification. Some
additional details of the trend of the curves should due to the
different responses of the horns to tensile stress.

So far, one finds that the deformed SWNH-120-SEG exhibits obvious
thermal rectification, and its dependence on strain seems similar to
the other two nanohorns in the paper. Thus we can take SWNH-120-SEG
as a typical example in the following text. In order to find the
maximum performance of SWNH-120-SEG, we adjust only the temperature
difference of two ends by fixing $L / L_0 $ to 1.015. The results
are presented in Fig. 3. For comparison, the temperature difference
dependence of undeformed SWNH-120-SEG is also shown.

In the inset of Fig. 3, the heat flux appears obviously asymmetric
in both structures, but the elongated structure shows larger
asymmetry. It can be found that the dependence of heat flux on
temperature difference rapidly deviate from linear relationship when
$\left| {\Delta T} \right|$ is larger than about $0.1$. And when
$\left| {\Delta T} \right|$ is larger than $0.5$, the rectification
can be as large as 2 in deformed SWNH-120-SEG. The overall maximum
rectification is about 2.4 when $\left| {\Delta T} \right| \approx
0.8$. This behavior seems very similar to the junction
situation,\cite{WG-LB2007} but the rectification in SWNH is more
obvious. This result is due to more distinct structural asymmetry in
SWNHs.

More significative thing is that a larger temperature difference can
induce a smaller heat flux when $|\Delta T| > 0.8$. This phenomenon
is called \textit{negative differential thermal resistance}
(NDTR)\cite{r6}, which is essential for thermal transistor. This
indicates that the SWNH might be used to construct thermal
transistors and other thermal devices such as thermal logic gates.

In summary, thermal rectification, i.e., asymmetric heat conduction,
has been studied in SWNHs by MD method. The structural asymmetry is
shown to have great effect on the asymmetric heat transport. The
horns with bigger top angle can show larger thermal rectification.
Our results also show that SWNHs behave like mass graded systems.

Furthermore, force constants can be adjusted by external stress to
influence the structure ununiformity. The competition between the
radius gradient and force constant gradient makes the system show
complicated response to the tensile strain. After choosing suitable
tensile strain, the thermal rectification in the SWNHs may become
more obvious than that in undeformed ones.


Our work suggests a possible efficient heat control device, and the
efficiency of this device can be controlled by simply applying
external stress. In this system, on-site potential is not requested
to achieve asymmetric heat conduction. Because SWNHs has been
fabricated experimentally, we believe the thermal controlling
devices can be made in the near future.

This work is supported by an Academic Research Grant, R-144-000-203-112,
 from the Ministry of Education of the Republic of
Singapore, and the DSTA under Project Agreement No. POD0410553.

\newpage

\begin{center}
{\bf Figures}
\end{center}

\begin{figure}
\includegraphics[width=\linewidth]{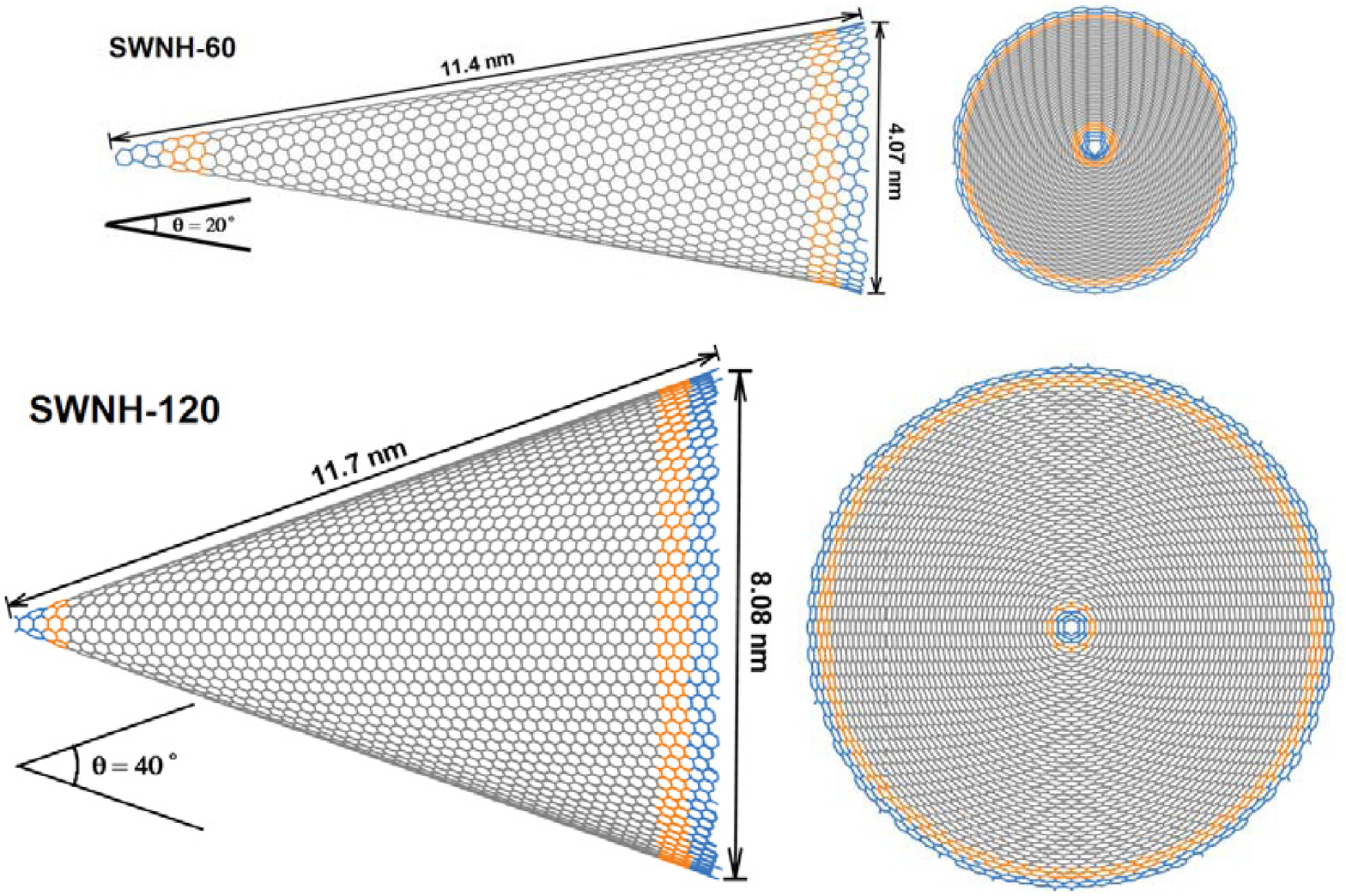}
\caption{(Color online). The structures of the two SWNHs in
our work. Upper panel shows the side view and top view of SWNH-60
structure. Lower panel shows the side view and top view of
SWNH-120 structure.}
\end{figure}

\begin{figure}
\includegraphics[width=\linewidth]{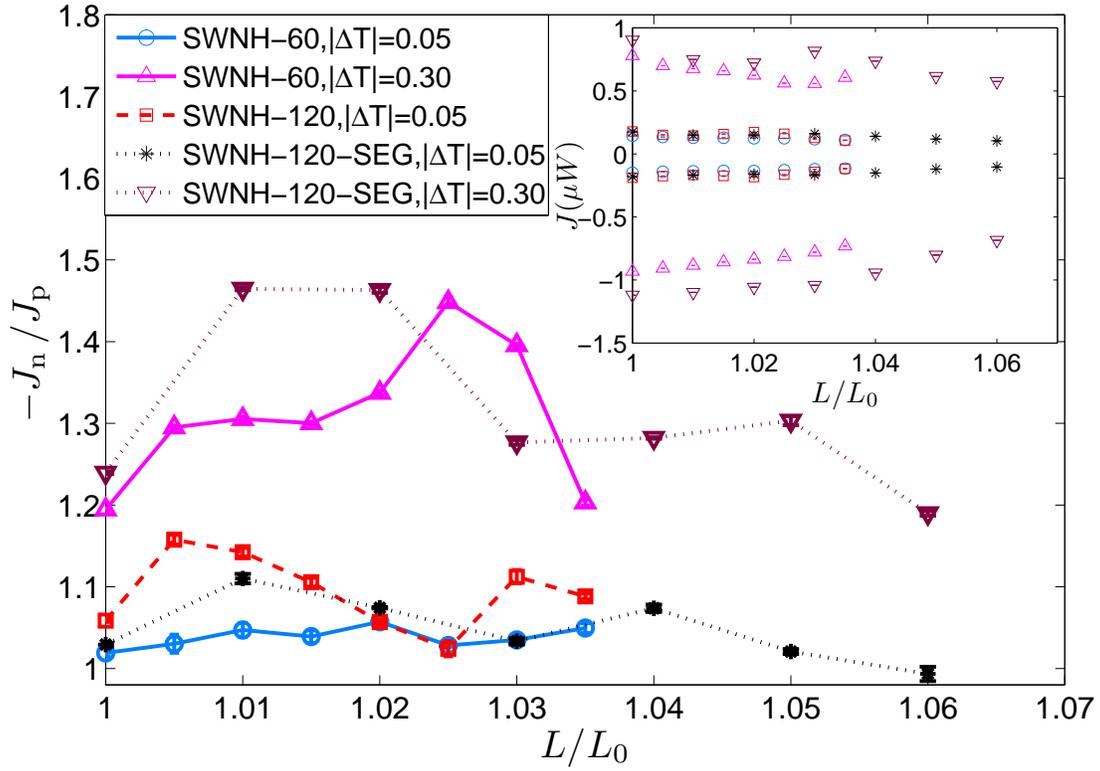}
\caption{(Color online). Thermal rectification in SWNHs
under different tensile strain versus top angle and temperature
difference. The negative heat fluxes correspond to negative
temperature differences $\Delta T$. Inset shows the corresponding
heat flux. Error bars are also plotted.}
\end{figure}


\begin{figure}
\includegraphics[width=\linewidth]{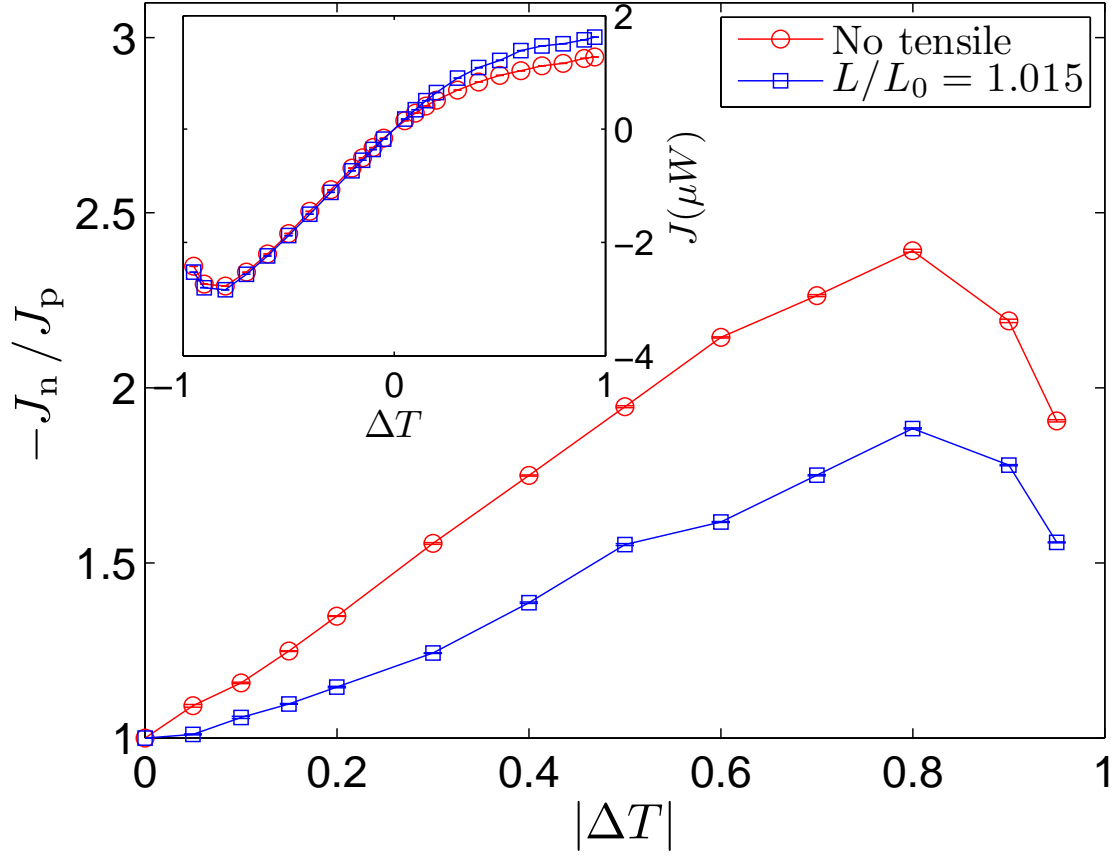}
\caption{(Color online). Thermal rectification of
SWNH-120-SEG under different temperature difference. Inset shows
the corresponding heat flux. Error bars are also plotted.}
\end{figure}

\end{document}